\documentclass[twocolumn,showpacs,preprintnumbers,amsmath,amssymb,nofootinbib]{revtex4}

\usepackage{graphicx}
\usepackage{color}
\usepackage{dcolumn}
\usepackage{bm}

\begin{document}


\title{Positioning in a flat two-dimensional space-time: the delay
master equation}

\author{Bartolom\'{e} Coll}
 \email{bartolome.coll@obspm.fr}
\author{Joan Josep Ferrando}
 \email{joan.ferrando@uv.es}
\author{Juan Antonio Morales-Lladosa}%
 \email{antonio.morales@uv.es}
\affiliation{%
Departament d'Astronomia i Astrof\'{\i}sica, \\Universitat de
Val\`encia, 46100 Burjassot, Val\`encia, Spain.
}%


\begin{abstract}
The basic theory on relativistic positioning systems in a
two-dimensional space-time has been presented in two previous papers
[Phys. Rev. D {\bf 73}, 084017 (2006); {\bf 74}, 104003 (2006)],
where the possibility of making relativistic gravimetry with these
systems has been analyzed by considering specific examples. Here we
study generic relativistic positioning systems in the Minkowski
plane. We analyze the information that can be obtained from the data
received by a user of the positioning system. We show that the
accelerations of the emitters and of the user along their
trajectories are determined by the sole knowledge of the emitter
positioning data and of the acceleration of only one of the
emitters. Moreover, as a consequence of the so called master delay
equation, the knowledge of this acceleration is only required during
an echo interval, {\em i.e.}, the interval between the
emission time of a signal by an emitter and its reception time after
being reflected by the other emitter. We illustrate these results
with the obtention of the dynamics of the emitters and of the user
from specific sets of data received by the user.
\end{abstract}

\pacs{04.20.-q, 95.10.Jk}

\maketitle

\section{Introduction}
\label{intro}

A {\em relativistic positioning system} is defined by four clocks
$\gamma_A$ ({\it emitters}) in arbitrary motion broadcasting their
proper times $\tau^A$ in some region of a (four-dimensional)
space-time \cite{coll-1,cfm-a,cfm-b,coll-pozo-1,Minko}. Then, every
event reached by the signals is naturally labeled by the four times
$\{\tau^A\}:$ the {\it emission coordinates} of this
event.\footnote{As a physical realization of a mathematical
coordinate system, the positioning system defined above presents
interesting qualities and, among them, those of being {\em generic},
(gravity-){\em free} and {\em immediate} \cite{cfm-a,
coll-1,coll-2,coll-3}.} The first to propose such physical
construction of emission coordinates seem to have been B. Coll
\cite{coll-2}. Up-to-date references on this concept and its
applications and a brief report on relativistic positioning can be
found in \cite{white}.

Although some explicit results have been obtained for generic
four-dimensional relativistic positioning \cite{coll-pozo-1,
Minko,pozo,bini,NosERE08,Pacome}, a full development of the theory
requires a previous training on simple and particular situations. A
two-dimensional approach to relativistic positioning systems allows
the use of precise and explicit diagrams which improve the
qualitative comprehension of general four-dimensional positioning
systems. The basic features of this two-dimensional approach and the
explicit relation between emission coordinates and any given null
coordinate system has been presented in \cite{cfm-a}. There, we have
also studied in detail the positioning system defined in flat
space-time by geodesic emitters.

In a subsequent work \cite{cfm-b} we have studied the possibility of
making relativistic gravimetry or, more generally, the possibility
of obtaining the dynamics of the emitters and/or of the user, as
well as the detection of the absence or presence of a gravitational
field and its measure. This possibility is examined by means of a
(non geodesic) {\em stationary positioning system} constructed in
two different scenarios: Minkowski and Schwarzschild planes.

In this work we go further in the analysis of two-dimensional
positioning problems. Until now \cite{cfm-b} we have considered
stationary or geodesic positioning systems in which the user had, a
priory, a partial or full information about the gravitational field
and a partial or full information about the positioning system. Here
we consider a new situation: the user knows the space-time where he
is immersed (flat, Schwarzschild,...) but he has no information
about the positioning system. Can the data received by the user
determine the characteristics of the positioning system? Can the
user obtain information on his local units of time and distance and
on his acceleration?

The answer to these questions is still an open problem for a generic
space-time, but in this work we undertake this query for Minkowski
plane and we analyze the minimum set of data that determine all the
user and system information. A remarkable result is that the data
received by a user of the positioning system are not independent
quantities because of they are submitted to what we call {\em the
public data constraints}. A consequence of these constraints is the
{\em delay master equation} which implies that the accelerations of
the emitters and of the user along their trajectories are determined
by the sole knowledge of the emitter positioning data and of the
acceleration of only one of the emitters and only during a (causal) {\em
echo interval}, {\em i.e.}, the interval between the
emission time of a signal by an emitter and its reception time after
being reflected by the other emitter.

In order to better understand our results we illustrate them with
two specific situations, the positioning systems defined,
respectively, by two inertial emitters or by two (stationary)
uniformly accelerated emitters. In them, starting from a partial set
of user data, we obtain the proper time and acceleration of the user
and we determine the full dynamical properties of the positioning
system.

The work is organized as follows. In Sec. \ref{section-II} we
summarize the basic concepts and notation about relativistic
positioning systems in a two-dimensional space-time. In Sec.
\ref{section-III} we obtain some constraint conditions which
restrict the user data and show that all the user and system
information can be obtained from the emitter positioning data and
the acceleration of only one of the emitters. Sec. \ref{section-IV}
and Sec. \ref{section-V} are devoted to illustrate these general
results by considering the above mentioned particular situations. In
Sec. \ref{section-VI} we deduce stronger restrictions on the user
data, the delay master equation, and we clarify the role that this
equation plays by applying it to the positioning systems considered
before. We finish in Sec. \ref{discussion} with a short discussion
about the present results and comments on prospective work.

A short communication of some results of this work was
presented in the Spanish Relativity meeting ERE-2007 \cite{ere2007}.

\begin{figure*}[htb]
    \includegraphics[angle=0,width=0.78\textwidth]{%
        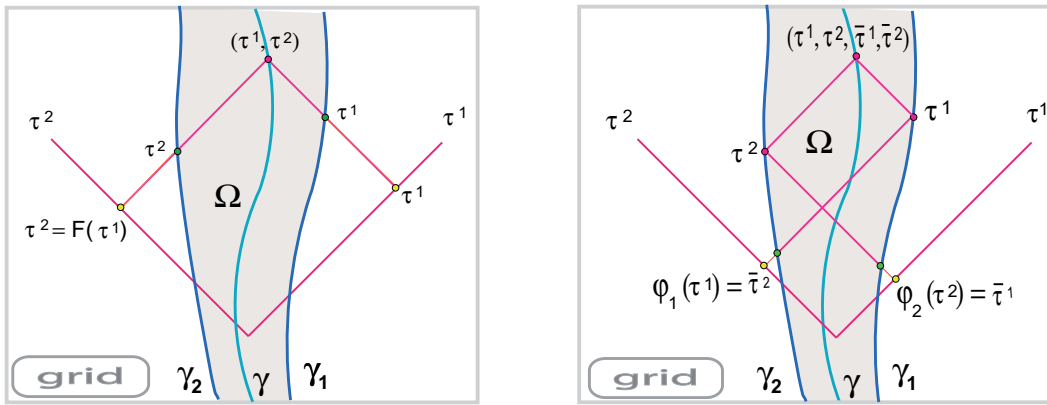}
    \caption{       \label{fig:grid-data}
            (a) Geometric interpretation of the emission
            coordinates: the proper times $\{\tau^1, \tau^2\}$
            received by a user $\gamma$ give his emission
            coordinates. These
            {\em user positioning data} $\{\tau^1, \tau^2\}$ allow
            the user to know his trajectory $\tau^2 = F(\tau^1)$
            in emission coordinates and he can draw it in the
            grid $\{\tau^1\}\times \{\tau^2\}$. (b)  The {\em
            emitter positioning data} $\{\tau^1, \tau^2; \bar{\tau}^1,
            \bar{\tau}^2\}$ allow the user to know the emitter
            trajectories $\varphi_1(\tau^1)$, $\varphi_2(\tau^2)$ in
            emission coordinates.}
\end{figure*}

\section{Two-dimensional approach}
\label{section-II}

In a two-dimensional space-time, a {\em relativistic positioning
system} is defined by two clocks, with world lines $\gamma_1$ and
$\gamma_2$ ({\em emitters}), broadcasting their proper times
$\tau^1$ and $\tau^2$ by mean of electromagnetic signals. In the
region $\Omega$ between both emitters, the past light cone of every
event cuts the emitter world lines at $\gamma_1(\tau^1)$ and
$\gamma_2(\tau^2)$, respectively. Then $\{\tau^1,\tau^2\}$ are the
{\em emission coordinates} of the event: the two proper time signals
received by any observer at the event from the two clocks (see Fig.
\ref{fig:grid-data}(a)). Nevertheless, the signals $\tau^1$ and
$\tau^2$ do not constitute coordinates for the events in the outside
region \cite{cfm-a}.

The plane $\{\tau^1\}\times\{\tau^2\}$ ($\tau^1, \tau^2 \in
\mathbb{R}$) in which the different data of the positioning system
can be transcribed is the {\em grid} of the positioning system. In
this grid, the trajectories of the two emitters define an interior
region and two exterior ones. This interior region in the grid is in
one-to-one correspondence with the interior region in the
space-time, i.e. with the set $\Omega$ of events that can be
distinguished by the pair of times $(\tau^1,\tau^2)$ that reach
them. But the exterior regions in the grid have no physical meaning
(see \cite{cfm-b} for more details on the grid).

An observer $\gamma,$ traveling throughout an emission coordinate
domain $\Omega$ and equipped with a receiver reading the received
proper times $(\tau^1,\tau^2)$ at each point of his trajectory, is
called a {\it user} of the positioning system.

We consider in this work {\em auto-locating positioning
systems}, which are systems in which every emitter clock not only
broadcasts its proper time but also the proper time that it receives
from the other. Thus, the physical components of an
auto-locating positioning system are \cite{cfm-a}:
\begin{description}
\item
a {\em spatial
segment} constituted by two emitters $\gamma_1$, $\gamma_2$
broadcasting their proper times $\tau^1,$ $\tau^2$ {\em and} the
proper times $\bar{\tau}^2$, $\bar{\tau}^1$ that they receive each
one from the other, and
\item
a {\em user segment} constituted by the set
of all users traveling in an internal domain $\Omega$ and receiving
these four broadcast times $\{\tau^1, \tau^2; \bar{\tau}^1,
\bar{\tau}^2\}$.
\end{description}

Any user receiving continuously the {\em user positioning data}
$\{\tau^1, \tau^2\}$ can extract the equation $F$ of his trajectory
in the grid (see Fig. \ref{fig:grid-data}(a)):
\begin{equation}  \label{user-trajectory}
\tau^2 = F(\tau^1) \, .
\end{equation}

On the other hand, any user receiving continuously the {\em emitter
positioning data} $\{\tau^1, \tau^2; \bar{\tau}^1, \bar{\tau}^2\}$
may extract from them not only the equation (\ref{user-trajectory})
of  his trajectory, but also the equations of the trajectories of
the emitters in the grid (see Fig. \ref{fig:grid-data}(b)):
\begin{equation} \label{emitter-trajectories}
\varphi_1(\tau^1) = \bar{\tau}^2 \, , \qquad  \varphi_2(\tau^2) =
\bar{\tau}^1 \, .
\end{equation}

Eventually, the emitters $\gamma_1$, $\gamma_2$ could carry
accelerometers and broadcast their acceleration $\alpha_1$,
$\alpha_2$, meanwhile the users $\gamma$ could be endowed with
receivers able to read the broadcast emitter accelerations
$\{\alpha_1, \alpha_2\}$. These new elements allow any user to know
the acceleration scalar of the emitters:
\begin{equation} \label{emitter-accelerations}
\alpha_1 = \alpha_1(\tau^1) \, , \qquad \alpha_2 = \alpha_2(\tau^2) \, .
\end{equation}

Users can also generate their own data, carrying a clock to measure
their proper time $\tau$ and/or an accelerometer to measure their
proper acceleration $\alpha$. The user's clock allows any user to
know his proper time function $\tau(\tau^1)$ (or $\tau(\tau^2)$)
and, consequently by using (\ref{user-trajectory}), to obtain the
proper time parametrization of his trajectory:
\begin{equation} \label{user-proper-time}
\gamma \equiv \{ \, \tau^1 = \psi^1(\tau) \, ; \ \tau^2 =
\psi^2(\tau)\} \, .
\end{equation}
The user's accelerometer allows any user to know his proper
acceleration scalar:
\[
\alpha = \alpha(\tau) \, .
\]

Thus, a relativistic positioning system may generate the {\it user
data}:
\begin{equation} \label{user-data}
\{\tau^1, \tau^2; \bar{\tau}^1, \bar{\tau}^2; \alpha_1, \alpha_2;
\tau, \alpha\} \, .
\end{equation}

The emitter trajectories (\ref{emitter-trajectories}) and the
emitter accelerations (\ref{emitter-accelerations}) do not depend
on the user that receives them. Thus, among the user data
(\ref{user-data}) we can distinguish the subsets:
\begin{itemize}
\item[(i)]
{\em emitter positioning data} $\{\tau^1, \tau^2; \bar{\tau}^1,
\bar{\tau}^2\}$,
\item[(ii)]
{\em public data} $\{\tau^1, \tau^2; \bar{\tau}^1, \bar{\tau}^2;
\alpha_1, \alpha_2\}$,
\item[(iii)]
{\em user proper data} $\{\tau,
\alpha\}$.
\end{itemize}

The purpose of the (relativistic) theory of positioning systems is
to develop the techniques necessary to determine the space-time
metric as well as the dynamics of emitters and users from (a subset
of) the user data.

In order to study specific positioning systems in known space-times,
it is useful to obtain the explicit expression of the emission
coordinates in terms of arbitrary null coordinates
$\{\texttt{u},\texttt{v}\}$.\footnote{In a two-dimensional
space-time,  null coordinates $\{\texttt{u},\texttt{v}\}$ are those
whose gradients, $du, \, dv$, determine light-like directions.} The
general method to obtain this transformation has been exposed in
\cite{cfm-a} and, in next section, we apply it to the inertial null
coordinates in flat space-time.\footnote{In a flat two-dimensional
space-time, for every inertial coordinate system $\{t,x\}$ we can
define the {\it inertial null coordinates}
$\{\texttt{u},\texttt{v}\}$: $\texttt{u} = t + x \, ,\ \texttt{v} =
t - x$. In this coordinates $\{\texttt{u},\texttt{v}\}$, the metric
tensor takes the form: $ \,d s^2 = dt^2 - d x^2 = d\texttt{u} \,
d\texttt{v}$.}

\section{Positioning in flat space-time}
\label{section-III}

In the development of the two-dimensional approach we have analyzed
situations \cite{cfm-a,cfm-b} under the assumption that the user
has a priory information about the positioning system, that is,
the user knows, at least partially, the dynamics of the emitters.
Now, we work under the weaker assumption that the user knows the
space-time where he is immersed but he has no a priory information about the
positioning system. Then, we want to analyze if the public data
received by the user afford information about: (i) his local unities
of time (ii) his acceleration, (iii) the metric in emission
coordinates, (iv) the coordinate transformation from emission
coordinates to a characteristic coordinate system of the given
space-time, and (v) his trajectory and emitter trajectories in this
characteristic coordinate system.

Although some results obtained elsewhere \cite{cfm-b} for the
Schwarzschild plane suggest that many of the results that we present
here could be generalized to non-flat space-times, from now on we
focus on the {\em flat case}.

\subsection{From emission to inertial coordinates}
\label{subsection-III-A}

Let us consider the positioning system defined by the emitters
$\gamma_1$ and $\gamma_2$ in the Minkowski plane, and let us assume
for the moment that the {\em proper time history of the emitters} is
known in an inertial null coordinate system
$\{\texttt{u},\texttt{v}\}$:
\begin{equation}  \label{emitters-inertial}
\gamma_1 \equiv \begin{cases} \texttt{u} = u_1(\tau^1) \\
\texttt{v} = v_1(\tau^1) \, ,\end{cases} \qquad \gamma_2 \equiv
\begin{cases} \texttt{u} = u_2(\tau^2) \\ \texttt{v} = v_2(\tau^2) \, .
\end{cases}
\end{equation}

The transformation from emission coordinates $\{\tau^1,\tau^2\}$ to
the inertial null system $\{\texttt{u},\texttt{v}\}$ is given by
\cite{cfm-a}:
\begin{equation}  \label{coordinatechange0}
\begin{array}{l}
\texttt{u} = u_1(\tau^1) \\
\texttt{v} = v_2(\tau^2) \, ,
\end{array}
\qquad \quad
\begin{array}{l}
\tau^1 = u_1^{-1}(\texttt{u}) = \tau^1(\texttt{u}) \\
\tau^2 = v_2^{-1}(\texttt{v}) = \tau^2(\texttt{v}) \, .
\end{array}
\end{equation}

Note that relations (\ref{coordinatechange0}) define {\it emission
coordinates} in the {\it emission coordinate domain} $\Omega$
between both emitters. But outside this region the transformation
(\ref{coordinatechange0}) also determines null coordinates, but they
are not emission coordinates for our positioning system, i.e. they
cannot be constructed by means of signals broadcasted by its two
clocks \cite{cfm-a}.

In emission coordinates, the emitter trajectories take the
expression:
\begin{equation}  \label{emitters-grid}
\gamma_1 \equiv \begin{cases} \tau^1 = \tau^1 \\ \tau^2 =
\varphi_1(\tau^1) \, ,\end{cases} \qquad \gamma_2 \equiv
\begin{cases} \tau^1 = \varphi_2(\tau^2) \\ \tau^2 = \tau^2 \, .
\end{cases}
\end{equation}
where, from (\ref{emitters-inertial}) and (\ref{coordinatechange0}),
the functions $\varphi_i$ are given by:
\begin{equation}  \label{phis}
\varphi_1 = v_2^{-1} \circ v_1    \, , \qquad   \varphi_2 = u_1^{-1}
\circ u_2  \, .
\end{equation}

Conversely, from this last formulas, we obtain:
\begin{equation}  \label{phis-tris}
v_1 = v_2 \circ \varphi_1    \, , \qquad \quad  u_2 = u_1 \circ
\varphi_2  \, .
\end{equation}
As obtained in (\ref{emitter-trajectories}), the emitter positioning
data determine the emitter trajectories $\varphi_i(\tau^i)$ in the
grid. Then, taking into account (\ref{emitters-inertial}) and the
expression of the transformation (\ref{coordinatechange0}),
relations (\ref{phis-tris}) give the precise expression of
the following simple fact:\\[0.5mm]
{\it Statement 1}.-- If one knows the transformation from emission
to inertial coordinates, the emitter positioning data $\{\tau^1,
\tau^2; \bar{\tau}^1, \bar{\tau}^2\}$ determine the proper time
history of the emitters in inertial coordinates.

\subsection{Metric in emission coordinates}
\label{subsection-III-B}

From the metric line element in inertial null coordinates
$\{\texttt{u},\texttt{v}\}$, $  \,d s^2 =  d\texttt{u} \,
d\texttt{v}$, and the coordinate transformation
(\ref{coordinatechange0}), we obtain that the metric tensor in
emission coordinates $\{\tau^1,\tau^2\}$ takes the expression:
\begin{equation} \label{metric-change}
\begin{array}{c}
ds^2 = m(\tau^1,\tau^2) d \tau^1 d\tau^2 \, , \\[2mm]
m(\tau^1,\tau^2) = u_1'(\tau^1) v_2'(\tau^2) \, .
\end{array}
\end{equation}

Can the functions $u_1(\tau^1)$ and $v_2(\tau^2)$ be determined from
the public data? Besides the emitter positioning data $\{\tau^1,
\tau^2; \bar{\tau}^1, \bar{\tau}^2\}$, the user needs dynamical
information of the system. Let us suppose, for the moment, that he
also receives the two emitter accelerations $\{\alpha_1,
\alpha_2\}$. Then, the acceleration scalar functions, $\alpha_i(\tau^i)$,
$i=1,2$, can be known from the public data, and the {\it emitter
shift parameters} $s_i$ can be calculated by means of (see
(\ref{A-shift})):
\begin{equation} \label{shift-alpha}
s_i(\tau^i) = \exp \left(\int \alpha_i(\tau^i) \, d \tau^i \right)
\, .
\end{equation}
Now we particularize the dynamic equation (\ref{A-dynamic-eq-flat})
for the the emitter $\gamma_1$ (resp., $\gamma_2$) by taking $\tau
=\tau^1$, $\psi_1(\tau^1) = \tau^1$, $\psi_2(\tau^1) =
\varphi_1(\tau^1)$ (resp., $\tau =\tau^2$, $\psi_1(\tau^2) =
\varphi_2(\tau^2)$, $\psi_2(\tau^2) = \tau^2$), and we obtain,
respectively:
\begin{equation} \label{dynamic-emitters-shift}
\begin{array}{c}
\displaystyle \,  s_1(\tau^1) = u_1'(\tau^1) =
\frac{1}{\dot{\varphi}_1(\tau^1)
v_2'(\varphi_1(\tau^1))} \, , \\[4mm]
\displaystyle s_2(\tau^2) = \frac{1}{v_2'(\tau^2)} =
\dot{\varphi}_2(\tau^2) u_1'(\varphi_2(\tau^2)) \, .
\end{array}
\end{equation}

Then, from these equations and expression (\ref{metric-change})
of the metric tensor, we obtain:\\[0.5mm]
{\it Statement 2}.-- In emission coordinates the metric function $m$
is given by the ratio between the shift of the emitters:
\begin{equation} \label{metric-shifts}
m(\tau^1, \tau^2) = \frac{s_1(\tau^1)}{s_2(\tau^2)} \, .
\end{equation}

Note that the user data determine every shift (\ref{shift-alpha}) up
to a constant factor which is related to the chosen inertial null
system $\{\texttt{u},\texttt{v}\}$. Of course, their ratio
(\ref{metric-shifts}) that gives the metric function in emission
coordinates does not depend on the inertial system. But, given the
emitter acceleration scalars, the constant factors which we take in
the two integrals (\ref{shift-alpha}) could correspond to two
different inertial systems. Nevertheless, we will see below that the
constraints on the public data allow to determine one emitter shift in
terms of the other emitter shift, both with respect the same
inertial system.

\subsection{Public data: constraint equations}
\label{subsection-III-C}

The emitter dynamic equations (\ref{dynamic-emitters-shift}) contain
essential information on the positioning system that we will now analyze.
From these four equalities we can eliminate $u_1'(\tau^1)$ and
$v_2'(\tau^2)$ and obtain the {\em constraint equations for the
emitter shifts}:
\begin{eqnarray}
s_2(\tau^2) = \dot{\varphi}_2(\tau^2) \, s_1(\varphi_2(\tau^2))
\label{12-shift-2} \, , \label{constraint-shift-1}
\\[1.5mm]
s_1(\tau^1) \, \dot{\varphi}_1(\tau^1) =  s_2(\varphi_1(\tau^1))
\label{12-shift-3}  \, . \label{constraint-shift-2}
\end{eqnarray}
Moreover, by differentiating with respect to the proper time one
obtains the {\em public data constraint equations}:
\begin{eqnarray}
\alpha_2(\tau^2) & = & \displaystyle
\frac{\ddot{\varphi}_2(\tau^2)}{\dot{\varphi}_2(\tau^2)} +
\dot{\varphi}_2(\tau^2)\, \alpha_1(\varphi_2(\tau^2))  \, ,
\label{constraint-acceleration-1}
\\[1.5mm]
\alpha_1(\tau^1) & = & \displaystyle -
\frac{\ddot{\varphi}_1(\tau^1)}{\dot{\varphi}_1(\tau^1)} +
\dot{\varphi}_1(\tau^1) \, \alpha_2(\varphi_1(\tau^1))   \, .
\label{constraint-acceleration-2}
\end{eqnarray}

Equations (\ref{constraint-acceleration-1}) and
(\ref{constraint-acceleration-2}) show that the public data
$\{\tau^1, \tau^2; \bar{\tau}^1, \bar{\tau}^2; \alpha_1, \alpha_2\}$
are not independent quantities. These constraints can be considered
as differential equations on the emitter trajectories
$\varphi_i(\tau^i)$ if the acceleration scalars $\alpha_i(\tau^i)$
are known, an approach that we will consider elsewhere. In the
present work we are interested in studying auto-locating positioning
systems for which the emitter positioning data $\{\tau^1, \tau^2;
\bar{\tau}^1, \bar{\tau}^2\}$ and, consequently, the functions
$\varphi_i(\tau^i)$ are known. From this point of view the public
data constraint equations (\ref{constraint-acceleration-1}) and
(\ref{constraint-acceleration-2}) state:\\[0.5mm]
{\it Statement 3}.-- If a user receives continuously the emitter
positioning data $\{\tau^1, \tau^2; \bar{\tau}^1, \bar{\tau}^2\}$
and only the acceleration of one of the emitters, then the user
knows the acceleration of the other emitter.

\subsection{Public data: metric and system information}
\label{subsection-III-D}

The constraint equations for the emitter shifts
(\ref{constraint-shift-1}) and (\ref{constraint-shift-2}) determine
the shift of an emitter with respect to an inertial system in terms
of the shift of the other emitter whit respect to the {\em same} inertial
system and the emitter positioning data $\{\tau^1, \tau^2;
\bar{\tau}^1, \bar{\tau}^2\}$. Then, as a consequence of statement
2,
we have:\\[0.5mm]
{\it Statement 4}.-- If a user receives continuously the emitter
positioning data $\{\tau^1, \tau^2; \bar{\tau}^1, \bar{\tau}^2\}$
and the acceleration of one of the emitters, then the user knows the metric
function $m(\tau^1, \tau^2)$ in emission coordinates.

On the other hand, if one knows the emitter shifts $s_1(\tau^1)$ and
$s_2(\tau^2)$ with respect to an inertial system then, as a consequence
of (\ref{dynamic-emitters-shift}), one knows the derivatives of the
transformation (\ref{coordinatechange0}) from emission to these
inertial null coordinates. Thus, we can obtain this transformation
up to two additive constants depending on the origin of the inertial
null system. Moreover, taking into account statement 1, we have:\\[0.5mm]
{\it Statement 5}.-- If a user receives continuously the emitter
positioning data $\{\tau^1, \tau^2; \bar{\tau}^1, \bar{\tau}^2\}$
and the acceleration of one of the emitters, then the user knows the
transformation from emission to inertial coordinates and the proper
time history of the emitters in inertial coordinates.

The analytic expression of the results in statements 3, 4 and 5
depends on which of the two accelerations is known. Now we explain
the steps to be followed to obtain all the system information when
the emitter positioning data $\{\tau^1, \tau^2; \bar{\tau}^1,
\bar{\tau}^2\}$ and one of the accelerations, say $\alpha_1$, are known.
\begin{description}
\item {\em Received user data:} $\{\tau^1, \tau^2; \bar{\tau}^1,
\bar{\tau}^2, \alpha_1\}$.
\item[] Step s1: From the pairs $\{\tau^1; \bar{\tau}^2\}$
and $\{\tau^2; \bar{\tau}^1\}$, determine the emitter trajectory
functions $\varphi_1(\tau^1)$ and $\varphi_2(\tau^2)$, respectively.
\item[] Step s2: From the pair $\{\tau^1; \alpha_1\}$,
determine the emitter acceleration scalar $\alpha_1(\tau^1)$.
\item[] Step s3: From the acceleration scalar $\alpha_1(\tau^1)$
obtained in step s2, determine the shift $s_1(\tau^1)$ with
respect to an inertial system $\{\texttt{u},\texttt{v}\}$:
    $$s_1(\tau^1) = \exp \left(\int \alpha_1(\tau^1)
\, d \tau^1 \right) \, .$$
\item[] Step s4: From the function $\varphi_2(\tau^2)$ obtained
in step s1 and the shift $s_1(\tau^1)$ obtained in step s3,
determine the shift $s_2(\tau^2)$ with respect to the inertial
system $\{\texttt{u},\texttt{v}\}$ and the acceleration scalar
$\alpha_2(\tau^2)$:
    $$\qquad s_2(\tau^2) = \dot{\varphi}_2(\tau^2) \,
s_1(\varphi_2(\tau^2))\, , \quad \alpha_2(\tau^2)  =  \displaystyle
\frac{\dot{s}_2(\tau_2)}{s_2(\tau_2)} \, .$$
\item[] Step s5: From the shifts $s_1(\tau^2)$ and $s_2(\tau^1)$
obtained in steps s3 and s4, determine the metric function in
emission coordinates:
    $$m(\tau^1, \tau^2) = \frac{s_1(\tau^1)}{s_2(\tau^2)} \, .$$
\item[] Step s6: From the shifts $s_1(\tau^1)$ and $s_2(\tau^2)$
obtained in steps s3 and s4, determine the transformation from
emission to inertial null coordinates $\{\texttt{u},\texttt{v}\}$:
        $$\begin{array}{l}
\displaystyle \texttt{u} = u_1(\tau^1) = \int s_1(\tau^1) \,
d \tau^1 \, ,\\[1.5mm]
\displaystyle \texttt{v} = v_2(\tau^2) = \int
\frac{1}{s_2(\tau^2)}\, d \tau^2  \, .
\end{array}$$
\item[] Step s7: From the functions $\varphi_1(\tau^1)$ and
$\varphi_2(\tau^2)$ obtained in step s1 and the coordinate
transformation $\{u_1(\tau^1), v_2(\tau^2)\}$ obtained in steps s6,
determine the proper time history of the emitters in inertial
null coordinates $\{\texttt{u},\texttt{v}\}$:
        $$\quad \quad \quad \gamma_1 \equiv \begin{cases} \texttt{u} =
u_1(\tau^1) \\ \texttt{v} = v_2(\varphi_1(\tau^1)) \, ,\end{cases}
\gamma_2 \equiv
        \begin{cases} \texttt{u} = u_1(\varphi_2(\tau^2)) \\
\texttt{v} = v_2(\tau^2)  \, . \end{cases}$$
\end{description}

Note that the shift $s_1(\tau^1)$ obtained in step s4 is fixed up to
a constant factor. Every choice of this constant determines a
different null inertial system $\{\texttt{u},\texttt{v}\}$ whose
origin depend on the choice of two additive constants when obtaining
$u_1(\tau^1)$ and $v_2(\tau^2)$ in step s6.

\subsection{Public data: user information}
\label{subsection-III-E}

Finally, we will see that the information provided by the proper
user data $\{\tau, \alpha\}$ can also be obtained from the emitter
positioning data $\{\tau^1, \tau^2; \bar{\tau}^1, \bar{\tau}^2\}$
and the acceleration of one emitter.

As explained in statement 4, the metric function in emission
coordinates can be obtained from these data. Moreover, from the user
positioning data  $\{\tau^1, \tau^2\}$ we can extract the trajectory
of the user in the grid, $\tau^2 = F(\tau^1)$. Then, the proper
time function $\tau(\tau^1)$ satisfies equation (\ref{A-trajectory})
which now becomes:
\begin{equation} \label{eq-proper-time}
[\tau'(\tau^1)]^2 = \frac{s_1(\tau^1)}{s_2(F(\tau^1))} F'(\tau^1) \,
.
\end{equation}

From the trajectory $\tau^2 = F(\tau^1)$ and the proper time
function $\tau(\tau^1)$ obtained from (\ref{eq-proper-time}), we can
get the proper time history of the user in emission coordinates,
$\tau^1 = \psi_1(\tau)$, $\tau^2 = \psi_2(\tau)$. Moreover, from
(\ref{A-dynamic-eq-flat}) and (\ref{A-shift}) we obtain the shift
and the acceleration of the user as:
\begin{equation} \label{eq-shift-user}
s(\tau) = \dot{\psi_1}(\tau) s_1(\psi_1(\tau)) \, , \quad
\alpha(\tau)= \frac{\dot{s}(\tau)}{s(\tau)} \, .
\end{equation}
As the coordinate transformation is also known (statement 5), we can
obtain the user proper time history in inertial coordinates.
Thus, we have:\\[0.5mm]
{\it Statement 6}.-- If a user receives the emitter positioning data
$\{\tau^1, \tau^2; \bar{\tau}^1, \bar{\tau}^2\}$ and the
acceleration of an emitter, then the user knows his local unities of
proper time, his acceleration and his proper time history in both
emission and inertial coordinates.

Equations (\ref{eq-proper-time}) and (\ref{eq-shift-user}) can be
useful in obtaining system information from the proper user data
$\{\tau, \alpha\}$, a question that we will consider elsewhere. Here
we suppose that the system information has been obtained, from the
emitter positioning data and one of the emitter accelerations, following the
steps s1-s7 presented in subsection above. Then, we can obtain the
user information enumerated in statement 6 in an alternative way
that is well adapted to the flat case. Indeed, from the user
trajectory in the grid and the coordinate transformation, we
determine the user trajectory in inertial null coordinates. Then, we
determine the proper time history in these coordinates, the user
shift and the scalar acceleration.

Now we explain the steps to be followed to obtain all these user
information when the emitter positioning data $\{\tau^1, \tau^2;
\bar{\tau}^1, \bar{\tau}^2\}$ and one of the accelerations, say $\alpha_1$,
are known.
\begin{description}
\item[] {\em Received user data:} $\{\tau^1, \tau^2; \bar{\tau}^1,
\bar{\tau}^2, \alpha_1\}$.
\item[] Step u1: From these data, and following steps s1,
s2, s3, s4 and s6, determine the coordinate transformation
$\{u_1(\tau^1), v_2(\tau^2)\}$ from emission to inertial coordinates
$\{\texttt{u},\texttt{v}\}$.
\item[] Step u2: From the pair $\{\tau^1; \tau^2\}$, determine
the user trajectory in the grid, $\tau^2 = F(\tau^1)$.
\item[] Step u3: From the user trajectory $\tau^2 = F(\tau^1)$
obtained in step u2 and the coordinate transformation
$\{u_1(\tau^1), v_2(\tau^2)\}$ obtained in step u1, determine the
user world line $\texttt{v}=f(\texttt{u})$ in the inertial system
$\{\texttt{u},\texttt{v}\}$:
    $$
    \texttt{v}=f(\texttt{u}) \, , \quad f = v_2 \circ F \circ u_1^{-1} \, .
    $$
\item[] Step u4: From the user world line
$\texttt{v}=f(\texttt{u})$ obtained in step u3, determine the
user proper time function $\tau = {\cal T}(\texttt{u})$:
    $$
    \tau = {\cal T}(\texttt{u})= \int \! \sqrt{f'(\texttt{u})}\,
    d  \texttt{u}  \, .
    $$
\item[] Step u5: From the user proper time function $\tau =
{\cal T}(\texttt{u})$ obtained in step u4 and the user world line
$\texttt{v}=f(\texttt{u})$ obtained in step u3, determine the proper
time history of the user in the inertial null coordinates
$\{\texttt{u},\texttt{v}\}$:
        $$\quad   \gamma \equiv
        \begin{cases} \texttt{u} = \texttt{u}(\tau) \, , \qquad
        {\cal T}(\texttt{u}(\tau))=\tau \\
        \texttt{v} = \texttt{v}(\tau) = f(\texttt{u}(\tau))  \, .
        \end{cases}
        $$
\item[] Step u6: From the proper time history of the user in the
inertial null coordinates $\{\texttt{u} = \texttt{u}(\tau),
\texttt{v} = \texttt{v}(\tau)\}$ obtained in step u5, determine
the shift $s(\tau)$ of the user with respect the inertial system
$\{\texttt{u},\texttt{v}\}$, and the user acceleration
$\alpha(\tau)$:
    $$s(\tau) = \dot{\texttt{u}}(\tau) \, , \qquad \quad
\alpha(\tau)= \frac{\ddot{\texttt{u}}(\tau)}{\dot{\texttt{u}}(\tau)} \, .
    $$
\item[] Step u7: From the proper time history of the user in the
inertial null coordinates $\{\texttt{u} = \texttt{u}(\tau),
\texttt{v} = \texttt{v}(\tau)\}$ obtained in step u5 and the
coordinate transformation $\{u_1(\tau^1), v_2(\tau^2)\}$ obtained in
step u1, determine the proper time history of the user in
emission coordinates:
    $$\quad \quad \gamma \equiv \begin{cases} \tau^1 = \psi_1(\tau) =
u_1^{-1}(\texttt{u}(\tau)) \, ,
\\[0.8mm]\tau^2 = \psi_2(\tau) = v_2^{-1}(\texttt{v}(\tau))
\, ,\end{cases}
    $$
and the proper time functions $\tau(\tau^1)$ and $\tau(\tau^2)$ of
the user:
$$
\quad \tau(\tau^1) = \psi_1^{-1}(\tau^1) \, , \quad \tau(\tau^2) =
\psi_2^{-1}(\tau^2) \, .
$$
\end{description}

Let us note that the proper time function obtained in step u4
depends on an additive constant which fixes the origin of the user
proper time.

\begin{figure*}[htb]
    \includegraphics[angle=0,width=0.78\textwidth]{%
        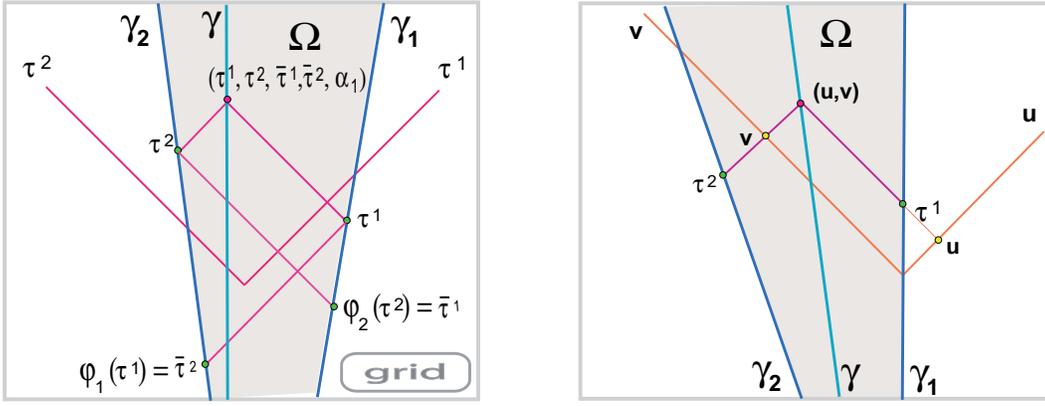}
    \caption{       \label{fig:geodesic-emitters}
            (a) {\em Emitter positioning data} $\{\tau^1, \tau^2;
            \bar{\tau}^1, \bar{\tau}^2\}$ allowing the user $\gamma$ to find
            that, in the grid, (i) the trajectories of the two emitters
            $\gamma_1$,  $\gamma_2$ are two straight lines with complementary
            slope, (ii) his own trajectory is a straight line parallel to the
            bisector. (b) If the user also receives an identically vanishing
            acceleration of an emitter, say $\alpha_1=0$, he obtains that
            he and the emitters have an inertial motion, and that his relative
            velocity with respect to every emitter is the same. Here we have
            drawn the trajectories in an inertial system at rest with respect
            to $\gamma_1$.}
\end{figure*}

\section{Information provided by the user data: the case of inertial
emitters} \label{section-IV}

The positioning system defined in Minkowski plane by two inertial
emitters has been analyzed in a previous paper \cite{cfm-a}. There
we started from the proper time history of the emitters in an
inertial null coordinate system and we studied what would be the
data that a user of the positioning system would receive. Here we
want to use this positioning system to illustrate the results
presented in the above section. Thus, now we will start, on one
hand, from the data received by an arbitrary user to obtain
information on the (positioning) system following the steps of
subsection \ref{subsection-III-D} and, on the other hand, from the
data received by a specific user to obtain information about himself
following the steps of subsection \ref{subsection-III-E}.

\subsection{System information} \label{subsection-IV-A}

\noindent {\em Assumption S}: The data $I \equiv \{\tau^1, \tau^2;
\bar{\tau}^1, \bar{\tau}^2, \alpha_1\}$ received by any user in the
emission coordinate domain is such that:
\begin{description}
\item[-] the pairs of data $\{\tau^1;
\bar{\tau}^2\}$ and $\{\tau^2; \bar{\tau}^1\}$ show a linear
relation with the same slope,
$$
\hspace*{3mm} \bar{\tau}^1 = \displaystyle  \tau^2_0 + \lambda \,
\tau^1 \, , \quad \bar{\tau}^2 = \displaystyle  \tau^1_0 + \lambda
\,\tau^2 \, ,
$$
{\em i.e.}, complementary slope in the grid $\{\tau^1, \tau^2\}$
(see Fig. \ref{fig:geodesic-emitters}(a)),
\item[-]
the acceleration $\alpha_1$ identically vanishes, $\alpha_1=0, \
\forall \, \tau^1$.
\end{description}

\begin{description}
\item[] Step s1: From the first item of this assumption {\em S}, any user
obtains that the emitter trajectory functions $\varphi_1(\tau^1)$
and $\varphi_2(\tau^2)$ are, respectively:
\begin{equation} \label{grid-inertial}
\hspace*{3mm} \varphi_1(\tau^1) = \displaystyle  \tau^2_0 + \lambda \, \tau^1 \, ,
\quad \varphi_2(\tau^2) = \displaystyle  \tau^1_0 + \lambda \,\tau^2
\, .
\end{equation}
\item[] Step s2: From the second item, any user
obtains that the emitter acceleration scalar $\alpha_1(\tau^1)$ is:
$$
\alpha_1(\tau^1) = 0 \, .
$$
\item[] Step s3: From the acceleration scalar $\alpha_1(\tau^1)$
obtained in step s2 any user obtains that the shift $s_1(\tau^1)$
with respect to any inertial system is constant. Let
$\{\texttt{u},\texttt{v}\}$ be an inertial system such that:
    $$s_1(\tau^1) = 1 \, .$$
\item[] Step s4: From the function $\varphi_2(\tau^2)$ obtained
in step s1 and the shift $s_1(\tau^1)$ obtained in step s3  any user
obtains that the shift $s_2(\tau^2)$ with respect to the inertial
system $\{\texttt{u},\texttt{v}\}$, and the acceleration
$\alpha_2(\tau^2)$ are, respectively:
$$\qquad s_2(\tau^2) = \lambda \, , \quad \alpha_2(\tau^2) = 0 \, .$$
\item[] Step s5: From the shifts $s_1(\tau^1)$ and $s_2(\tau^2)$
obtained in steps s3 and s4  any user obtains that the metric
function in emission coordinates is:
    $$m(\tau^1, \tau^2) = \frac{1}{\lambda} \, .$$
\item[] Step s6: From the shifts $s_1(\tau^1)$ and $s_2(\tau^2)$
obtained in steps s3 and s4 any user obtains that the
transformation from emission to the inertial null system
$\{\texttt{u},\texttt{v}\}$ (for a choice of the origin) is:
        $$\begin{array}{l}
\displaystyle \texttt{u} = u_1(\tau^1) = \tau^1  \, ,\\[1.5mm]
\displaystyle \texttt{v} = v_2(\tau^2) = \frac{1}{\lambda}(\tau^2 -
\tau^2_0) \, .
\end{array}$$
\item[] Step s7: From the functions $\varphi_1(\tau^1)$ and
$\varphi_2(\tau^2)$ obtained in step s1 and the coordinate
transformation $\{u_1(\tau^1), v_2(\tau^2)\}$ obtained in step s6
any user obtains that the proper time history of the emitters in the
inertial null coordinates $\{\texttt{u},\texttt{v}\}$ are, respectively:
        $$\quad \quad \gamma_1 \equiv \begin{cases} \texttt{u} =
\tau^1 \\ \texttt{v} = \tau^1 \, , \end{cases} \quad \gamma_2 \equiv
        \begin{cases} \texttt{u} = \tau^1_0 + \lambda \,\tau^2 \\
\texttt{v} = \frac{1}{\lambda}(\tau^2 - \tau^2_0)  \, .
\end{cases}$$
\end{description}

Steps s2 and s4 show that {\em a user can receive the assumed set of data $I$
only if the positioning system is defined by two inertial emitters}.
In step 3, the arbitrary constant factor has been chosen so that
emitter $\gamma_1$ is at rest with respect the inertial system
$\{\texttt{u},\texttt{v}\}$ (see Fig.
\ref{fig:geodesic-emitters}(b)). Moreover, from step s6 we obtain
that, in the orthonormal coordinate system $\{t,x\}$ associated with
the null one $\{\texttt{u},\texttt{v}\}$, the proper time history of
the emitter $\gamma_1$ is $\{t=\tau^1; \ x=0\}$. This means that we
have chosen the additive constants in step s6 so that the origin of
the inertial system is at the event which the emitter $\gamma_1$
reaches when his proper time clock watches zero.

\subsection{User information} \label{subsection-IV-B}

Now we will illustrate how a specific user, receiving the emitter
positioning data and the acceleration of one of the emitters, can
determine his time and his dynamics.

\vspace{2mm}
\noindent {\em Assumption U}: The specific user in question receives
the user data $I \equiv \{\tau^1, \tau^2; \bar{\tau}^1,
\bar{\tau}^2, \alpha_1\}$ of the above assumption {\em S} and, in
addition:
\begin{description}
\item[-] the data $\{\tau^1; \tau^2\}$ show a linear relation with
slope $1$ (see Fig. \ref{fig:geodesic-emitters}(a)).
\end{description}

\begin{description}
\item[] Step u1: From these data, and following steps s1,
s2, s3, s4 and s6 above, the user has obtained the coordinate
transformation $\{u_1(\tau^1), v_2(\tau^2)\}$ from emission to
inertial coordinates $\{\texttt{u},\texttt{v}\}$.
\item[] Step u2: From the above assumption {\em U} the user obtains that
his trajectory in the grid is:
    $$\tau^2 = F(\tau^1) = \tau^1 + C \, .$$
\item[] Step u3: From this user trajectory $\tau^2 = F(\tau^1)$
obtained in step u2 and the coordinate transformation
$\{u_1(\tau^1), v_2(\tau^2)\}$ obtained in step u1 the user obtains that
his world line $\texttt{v}=f(\texttt{u})$ in the inertial
system $\{\texttt{u},\texttt{v}\}$ is:
    $$
    \texttt{v}= f(\texttt{u}) = \frac{1}{\lambda}(\texttt{u} +
    C - \tau_0^2) \, .
    $$
\item[] Step u4: From the user world line
$\texttt{v}=f(\texttt{u})$ obtained in step u3 the user can obtain
that his proper time function $\tau = {\cal T}(\texttt{u})$ is:
    $$
    \tau = {\cal T}(\texttt{u})= \frac{1}{\sqrt{\lambda}}\, \texttt{u}  \, .
    $$
\item[] Step u5: From the user proper time function $\tau =
{\cal T}(\texttt{u})$ obtained in step u4 and the user world line
$\texttt{v}=f(\texttt{u})$ obtained in step u3 the user obtains that
his proper time history in the inertial null coordinates
$\{\texttt{u},\texttt{v}\}$ is:
        $$\quad   \gamma \equiv
        \begin{cases} \texttt{u} = \texttt{u}(\tau) = \sqrt{\lambda}\,
        \tau \\ \displaystyle
        \texttt{v} = \texttt{v}(\tau) = \frac{1}{\sqrt{\lambda}}\, \tau +
        \frac{1}{\lambda}(C- \tau_0^2)  \, . \end{cases}$$
\item[] Step u6: From the proper time history of the user in the
inertial null coordinates $\{\texttt{u} = \texttt{u}(\tau),
\texttt{v} = \texttt{v}(\tau)\}$ obtained in step u5 the user obtains that
his shift $s(\tau)$ with respect the inertial system
$\{\texttt{u},\texttt{v}\}$, and his acceleration
$\alpha(\tau)$:
    $$s(\tau) = \sqrt{\lambda} \, , \qquad \quad \alpha(\tau)= 0 \, .
    $$
\item[] Step u7: From the proper time history of the user in the
inertial null coordinates $\{\texttt{u} = \texttt{u}(\tau),
\texttt{v} = \texttt{v}(\tau)\}$ obtained in step u5 and the
coordinate transformation $\{u_1(\tau^1), v_2(\tau^2)\}$ obtained in
step u1 the user obtains that his proper time history in
emission coordinates is:
    $$\quad \quad \quad \gamma \equiv \begin{cases} \tau^1 = \psi_1(\tau) =
    \sqrt{\lambda} \, \tau  \, , \\[0.8mm]
    \tau^2 = \psi_2(\tau) = \sqrt{\lambda} \, \tau + C \, . \end{cases}
    $$
and the user proper time lapse $\Delta \tau$ is:
$$
\Delta \tau = \frac{1}{\sqrt{\lambda}}\, \Delta \tau^1 =
\frac{1}{\sqrt{\lambda}}\, \Delta \tau^2 \, .
$$
\end{description}

Let us note that the hyperbolic angle between the trajectories of
the user and the emitter $\gamma_1$ is $\phi = \ln s(\tau) = \frac12
\ln \lambda$, and the  hyperbolic angle between the trajectories of
the emitters $\gamma_2$ and $\gamma_1$ is $\phi_2 = \ln s_2(\tau^2)
= \ln \lambda = 2 \phi$. Consequently, the user has the same
relative velocity with respect to both emitters. (see Fig.
\ref{fig:geodesic-emitters}(b)). On the other hand, in the proper
time function obtained in step u4 we have chosen the additive
constant so that the user proper time clock watches zero when time
$\tau^1 = 0$ is received by the user.

\section{Information provided by the user data: the case of
stationary emitters} \label{section-V}

The positioning system defined in Minkowski plane by two
(stationary) uniformly accelerated emitters has been analyzed in a
previous paper \cite{cfm-b}. There we supposed that the user knew, a
priory, that the system was stationary. Here we start from the
emitter positioning data and the acceleration of an emitter and,
following the steps presented in subsections \ref{subsection-III-D}
and \ref{subsection-III-E}, we obtain all the system and user
information.

\subsection{System information} \label{subsection-V-A}

\begin{figure*}[htb]
    \includegraphics[angle=0,width=0.76\textwidth]{%
        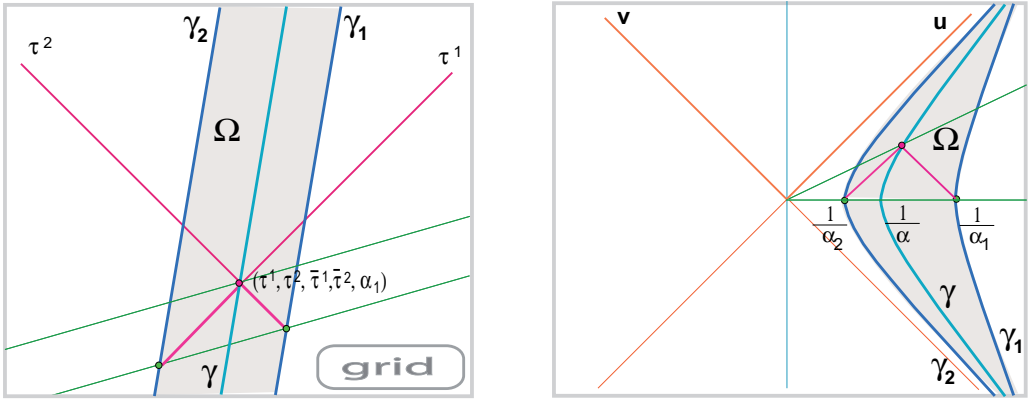}
    \caption{       \label{fig:accelerated}
            (a) {\em Emitter positioning data} $\{\tau^1,
            \tau^2; \bar{\tau}^1, \bar{\tau}^2\}$ allowing the user
            $\gamma$ to find that, in the grid, (i) the trajectories
            of the two emitters $\gamma_1$,  $\gamma_2$ are two parallel
            straight lines, (ii) his own trajectory is a straight line
            parallel to the emitters. Here we have plotted the case
            $c=0$ and we have stressed the user when receiving vanishing
            emitter coordinates. (b) If the user also receives the
            acceleration of the emitter $\gamma_1$ with the constant
            value $\alpha_1=\frac{1}{q}\ln \omega$, where $\omega$ is the
            {\em slope parameter} and $q$ is the {\em separation parameter},
            he obtains that he and the emitters have a non
            inertial stationary motion, and he can determine their
            constant accelerations and their synchronization. Here we
            have drawn the trajectories when the {\em synchronization
            parameter} $\sigma=0$. In green we have drawn the locus of
            simultaneous events for the stationary congruence.}
\end{figure*}

\noindent {\em Assumption S}: The data $A \equiv \{\tau^1, \tau^2;
\bar{\tau}^1, \bar{\tau}^2, \alpha_1\}$ received by any user in the
emission coordinate domain is such that:
\begin{description}
\item[-] the pairs of data $\{\tau^1;
\bar{\tau}^2\}$ and $\{\tau^2; \bar{\tau}^1\}$ show
linear relations with inverse slopes,
$$
\hspace*{3mm} \bar{\tau}^1 = \displaystyle  \frac{1}{\omega}(\tau^1
- q - \sigma)  \, , \quad \bar{\tau}^2 = \displaystyle  \omega
\tau^2 - q + \sigma \,  ,
$$
with $\omega>1$ and $q>0$, {\em i.e.}, parallel straight lines in
the grid $\{\tau^1, \tau^2\}$ (see Fig. \ref{fig:accelerated}(a)),
\item[-]
the acceleration
$\alpha_1$ takes the constant value $\alpha_1=\frac{1}{q}\ln
\omega , \ \forall \, \tau^1$.
\end{description}

\begin{description}
\item[] Step s1: From the first item of this assumption {\em S}, any user
obtains that the emitter
trajectory functions $\varphi_1(\tau^1)$ and $\varphi_2(\tau^2)$
are, respectively:
\begin{equation} \label{grid-stationary}
\begin{array}{l}
\varphi_1(\tau^1) = \displaystyle   \frac{1}{\omega}(\tau^1
- q - \sigma) \, , \\[3mm] \varphi_2(\tau^2) = \displaystyle  \omega
\tau^2 - q + \sigma \, .
\end{array}
\end{equation}
\item[] Step s2: From the second item any user
obtains that the emitter acceleration scalar $\alpha_1(\tau^1)$
is:
$$
\alpha_1(\tau^1) = \frac{1}{q} \ln \omega \equiv \alpha_1 \, .
$$
\item[] Step s3: From the acceleration scalar $\alpha_1(\tau^1)$
obtained in step s2 any user obtains that the shift $s_1(\tau^1)$
with respect to an inertial system $\{\texttt{u},\texttt{v}\}$
(fixed up to a choice of the origin) is:
    $$s_1(\tau^1) = \exp(\alpha_1 \tau^1) \, .$$
\item[] Step s4: From the function $\varphi_2(\tau^2)$ obtained
in step s1 and the shift $s_1(\tau^1)$ obtained in step s3 any user
obtains that the shift $s_2(\tau^2)$ with respect to the inertial system
$\{\texttt{u},\texttt{v}\}$, and the acceleration $\alpha_2(\tau^2)$
are:
$$\qquad s_2(\tau^2) = \exp(\alpha_2(\tau^2 - \tau^2_0)) \, ,
\quad \alpha_2(\tau^2) = \alpha_2 \, ,$$ where $\alpha_2 \equiv
\omega \alpha_1$, and $\tau^2_0 \equiv - \frac{\sigma}{\omega}$.
\item[] Step s5: From the shifts $s_1(\tau^1)$ and $s_2(\tau^1)$
obtained in steps s3 and s4 any user obtains that the metric function in
emission coordinates is:
    $$\displaystyle m(\tau^1, \tau^2) = \omega^{\frac{1}{q}(\tau^1- \omega
    \tau^2-\sigma)} \, .$$
\item[] Step s6: From the shifts $s_1(\tau^1)$ and $s_2(\tau^2)$
obtained in steps s3 and s4 any user obtains that the transformation
from emission to the inertial null system
$\{\texttt{u},\texttt{v}\}$ (for a choice of the origin) is:
        $$\begin{array}{l}
\displaystyle \texttt{u} = u_1(\tau^1) = \frac{1}{\alpha_1}\,
\exp(\alpha_1 \tau^1)  \, ,\\[3mm]
\displaystyle \texttt{v} = v_2(\tau^2) = - \frac{1}{\alpha_2}\,
\exp(-\alpha_2(\tau^2 - \tau^2_0)) \, .
\end{array}$$
\item[] Step s7: From the functions $\varphi_1(\tau^1)$ and
$\varphi_2(\tau^2)$ obtained in step s1 and the coordinate
transformation $\{u_1(\tau^1), v_2(\tau^2)\}$ obtained in step s6 any user
obtains that the proper time history of the emitters in inertial null
coordinates $\{\texttt{u},\texttt{v}\}$ is:
        $$\begin{array}{l}\quad \gamma_1\! \equiv\! \begin{cases}
\displaystyle \texttt{u} = \frac{1}{\alpha_1}\, \exp(\alpha_1
\tau^1) \\ \displaystyle  \texttt{v} =
-\frac{1}{\alpha_1}\, \exp(-\alpha_1 \tau^1) \, , \end{cases} \\[7mm] \quad
\gamma_2\! \equiv \! \begin{cases} \displaystyle  \texttt{u} =
\frac{1}{\alpha_2}\, \exp(\alpha_2(\tau^2 - \tau^2_0))
\\\displaystyle \texttt{v} = -\frac{1}{\alpha_2}\,
\exp(-\alpha_2(\tau^2 - \tau^2_0)) \, . \end{cases}
\end{array}$$
\end{description}

Steps s2 and s4 show that {\em a user receiving the set of data $A$
is, necessarily, in the coordinate domain of a positioning system
defined by two uniformly accelerated emitters with constant
acceleration scalars $\alpha_1(\tau_1)= \frac{1}{q} \ln \omega
\equiv \alpha_1$ and $\alpha_2(\tau_2) =\omega \alpha_1> \alpha_1$}.
In step s3, the arbitrary constant factor has been chosen so that
emitter $\gamma_1$ is at rest with respect the inertial system
$\{\texttt{u},\texttt{v}\}$ when his proper time clock watches zero.

From step s7, we have that the emitter trajectories in the inertial
system are $\alpha_i^2 \texttt{u} \texttt{v} = -1$. This means that
in step s6 we could choose the additive constants ({\em i.e.}, the
origin of the inertial coordinate system) so that the coordinate
bisectors are the asymptotes of both emitter trajectories (see Fig.
\ref{fig:accelerated}(b)). Thus, the emitters maintain a constant
radar distance and, consequently, they belong to a congruence of
stationary observers. On the other hand, $\tau^2_0 \equiv -
\frac{\sigma}{\omega}$ gives the time which watches the proper time
clock of $\gamma_2$ at the event simultaneous to the event where the
proper time clock of $\gamma_1$ watches zero. This fact shows that
in relativistic positioning the synchronization between the emitter
clocks is not necessary, but it can be extracted from the emitter
data.

\begin{figure*}[htb]
    \includegraphics[angle=0,width=0.77\textwidth]{%
    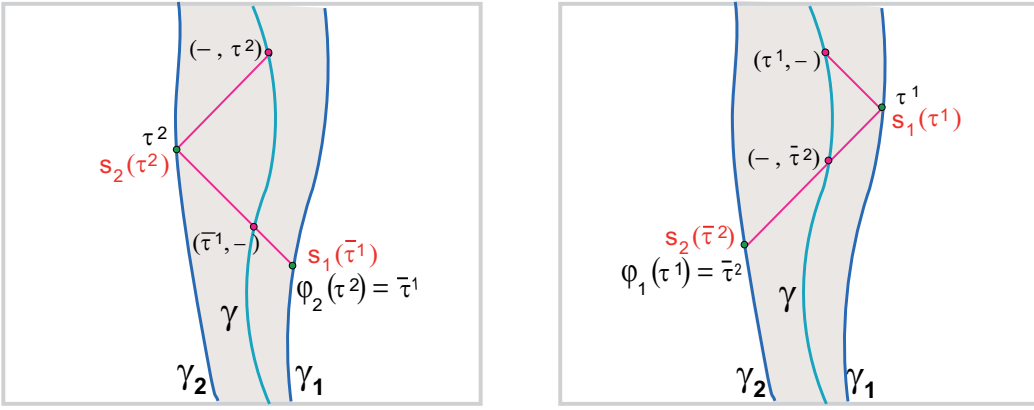}
    \caption{       \label{fig:constraints}
            Geometric interpretation of the constraint equations: (a) If
            a user receives the trajectory $\bar{\tau}^1 =
            \varphi_2(\tau^2)$ in the vicinity of time $\tau^2$ and the
            shift $s_1(\bar{\tau}^1)$ at time $\bar{\tau}^1$, then he
            can obtain the shift $s_2(\tau^2)$ at time $\tau^2$. (b)
            If a user receives the trajectory $\bar{\tau}^2 =
            \varphi_1(\tau^1)$ in the vicinity of time $\tau^1$ and the
            shift $s_2(\bar{\tau}^2)$ at time $\bar{\tau}^2$, then he
            can obtain the shift $s_1(\tau^1)$ at time $\tau^1$.}
\end{figure*}

\subsection{User information} \label{subsection-V-B}

Now we will illustrate how a specific user, receiving the emitter
positioning data and the acceleration of one of the emitters, can determine
his time and his dynamics.

\vspace{2mm}
\noindent {\em Assumption U}: The specific user in question receives
the user data $A \equiv \{\tau^1, \tau^2; \bar{\tau}^1,
\bar{\tau}^2, \alpha_1\}$ of the above assumption {\em S} and, in
addition:
\begin{description}
\item[-] the data $\{\tau^1; \tau^2\}$ show a linear relation with the
same slope than the emitters (parallel to the emitter trajectories
in the grid  $\{\tau^1, \tau^2\}$; see Fig.
\ref{fig:accelerated}(a)).
\end{description}

\begin{description}
\item[] Step u1: From these data, and following steps s1,
s2, s3, s4 and s6 above, the user has obtained the coordinate
transformation $\{u_1(\tau^1), v_2(\tau^2)\}$ from emission to
inertial coordinates $\{\texttt{u},\texttt{v}\}$.
\item[] Step u2: From the above assumption {\em U} the user can
obtain that his trajectory in the grid is:
    $$\qquad \tau^2 = F(\tau^1) =  \frac{1}{\omega} (\tau^1 -c)\, ,
    \quad q < c - \sigma < q \, .$$
\item[] Step u3: From the user trajectory $\tau^2 = F(\tau^1)$
obtained in step u2 and the coordinate transformation
$\{u_1(\tau^1), v_2(\tau^2)\}$ obtained in step u1 the user obtains
that his world line $\texttt{v}=f(\texttt{u})$ in the inertial
system $\{\texttt{u},\texttt{v}\}$ is:
    $$ \qquad
    \begin{array}{l}
    \displaystyle \texttt{v} = f(\texttt{u}) = - \frac{1}{\alpha^2
    \texttt{u}}\, , \qquad  \texttt{u} \, \texttt{v} = -
    \frac{1}{\alpha^2} \, ,\\[3mm]
    \displaystyle \alpha \equiv \frac{\ln \omega}{q}
    \omega^{\frac{1}{2q}(q+\sigma-c)} =
    \alpha_1^{\frac{1}{2q}(q-\sigma+c)}\alpha_2^{\frac{1}{2q}(q+\sigma-c)}
     \, .
     \end{array}
    $$
\item[] Step u4: From the user world line
$\texttt{v}=f(\texttt{u})$ obtained in step u3 the user obtains that
his proper time function $\tau = {\cal T}(\texttt{u})$ is:
    $$
    \tau = {\cal T}(\texttt{u})= \frac{1}{\alpha}\, \ln (\alpha \texttt{u})  \, .
    $$
\item[] Step u5: From the user proper time function $\tau =
{\cal T}((\texttt{u})$ obtained in step u4 and the user world line
$\texttt{v}=f(\texttt{u})$ obtained in step u3 the user obtains that
his proper time history in the inertial null coordinates
$\{\texttt{u},\texttt{v}\}$ is:
        $$\quad   \gamma \equiv
        \begin{cases} \displaystyle \texttt{u} = \texttt{u}(\tau) =
        \frac{1}{\alpha}\, \exp(\alpha \tau) \\ \displaystyle
        \texttt{v} = \texttt{v}(\tau) = -\frac{1}{\alpha}\,
        \exp(-\alpha \tau)  \, . \end{cases}$$
\item[] Step u6: From the proper time history of the user in the
inertial null coordinates $\{\texttt{u} = \texttt{u}(\tau),
\texttt{v} = \texttt{v}(\tau)\}$ obtained in step u5 the user obtains that
his shift $s(\tau)$ with respect the inertial system
$\{\texttt{u},\texttt{v}\}$, and his acceleration
$\alpha(\tau)$ are, respectively:
    $$s(\tau) = \exp(\alpha \tau) \, , \qquad \quad \alpha(\tau)= \alpha \, .
    $$
\item[] Step u7: From the proper time history of the user in the
inertial null coordinates $\{\texttt{u} = \texttt{u}(\tau),
\texttt{v} = \texttt{v}(\tau)\}$ obtained in step u5 and the
coordinate transformation $\{u_1(\tau^1), v_2(\tau^2)\}$ obtained in
step u1 the user obtains that his proper time history in
emission coordinates is:
    $$\quad \quad \gamma \equiv \begin{cases} \displaystyle \tau^1 =
    \frac{\alpha}{\alpha_1}  \, \tau - \frac{1}{2}(q+\sigma-c)  \, ,
    \\[2mm]  \displaystyle \tau^2 = \frac{\alpha}{\alpha_2} \, \tau
    - \frac{1}{2\omega}(q+\sigma+c)  \, . \end{cases}
    $$
and his proper time lapse $\Delta \tau$ is:
$$
\Delta \tau = \frac{\alpha_1}{\alpha}\, \Delta \tau^1 =
\frac{\alpha_2}{\alpha}\, \Delta \tau^2 \, ,
$$
where $\frac{\alpha}{\alpha_1} \equiv \omega^{\frac{1}{2q}(q+\sigma-c)}$
and $\frac{\alpha}{\alpha_2} \equiv \omega^{-\frac{1}{2q}(q-\sigma+c)}$.
\end{description}

Step u3 shows that the user also follows a stationary motion that
keep a constant radar distance with respect the two emitters (see Fig.
\ref{fig:accelerated}(b)).
Moreover, the constant value of the acceleration of the user is the
weighted geometric mean of the emitters' accelerations. In the
proper time function obtained in step u4 we have chosen the additive
constant so that the events, where the proper time clocks of the
user and of the emitter $\gamma_1$ watch zero, are simultaneous.

\section{The delay master equation}
\label{section-VI}

In Sec. \ref{section-III} we have shown that, as a consequence of
the public data constraint equations
(\ref{constraint-acceleration-1}) and
(\ref{constraint-acceleration-2}), the emitter positioning data an
the acceleration of an emitter determine the acceleration of the
other emitter. Nevertheless, in the steps given in subsections
\ref{subsection-III-D} and \ref{subsection-III-E}, which allow to
obtain all the system and user information, we only used one of
these two restrictions or, more precisely only one of the two
constraint equations for the shift (\ref{constraint-shift-1}) and
(\ref{constraint-shift-2}). Do these equations impose stronger
restrictions on the public data?
\begin{figure*}[htb]
    \includegraphics[angle=0,width=0.77\textwidth]{%
    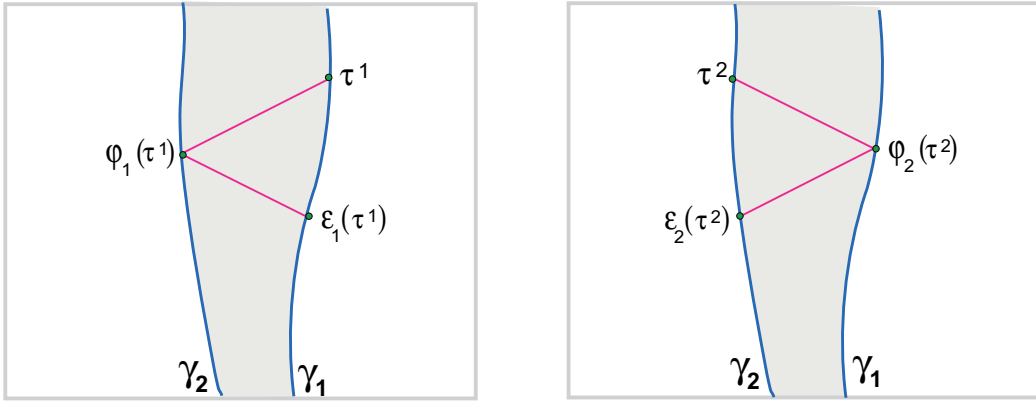}
    \caption{       \label{fig:echo}
            Geometric interpretation of the past echo functions
            $\varepsilon_i$ and the echo intervals
            $[\varepsilon_1(\tau^1), \tau^1]$ and
            $[\varepsilon_2(\tau^2), \tau^2]$: (a) If $\gamma_1$
            receives at time $\tau^1$ a signal after being echoed
            by $\gamma_2$, it must be emitted at time
            $\varepsilon_1(\tau^1)$. (b) If $\gamma_2$ receives
            at time $\tau^2$ a signal after
            being echoed by $\gamma_1$, it must be emitted at time
            $\varepsilon_2(\tau^2)$.}
\end{figure*}

In this section we will see that the answer is affirmative by
obtaining the precise restrictions that the emitter positioning data
impose on the dynamics of the emitters. This study requires to
consider the shift constraint equations (\ref{constraint-shift-1})
and (\ref{constraint-shift-2}), not as two independent equations,
but as a {\em constraint system}:
\begin{eqnarray}
s_2(\tau^2) = \dot{\varphi}_2(\tau^2) \, s_1(\varphi_2(\tau^2))
\label{12-shift-2} \, , \label{constraint-shift-1-bis}
\\[1.5mm]
s_1(\tau^1) \, \dot{\varphi}_1(\tau^1) =  s_2(\varphi_1(\tau^1))
\label{12-shift-3}  \, . \label{constraint-shift-2-bis}
\end{eqnarray}

\subsection{The (past) echo functions and the delay master equation}
\label{subsec-VI-A}

In Secs. \ref{section-III},  \ref{section-IV} and  \ref{section-V},
when we obtained an emitter acceleration from the emitter positioning
data and the acceleration of the other emitter, we supposed
that the user received continuously these data. Now, in order to
better understand the constraints on the public data, it is useful
to analyze its local behavior. In this sense, the constraint system
(\ref{constraint-shift-1-bis})-(\ref{constraint-shift-2-bis})
can be read as follows (see Fig. \ref{fig:constraints}):\\[0.5mm]
{\it Statement 7}.-- (i) If a user receives the trajectory
$\bar{\tau}^1 = \varphi_2(\tau^2)$ in the vicinity of time $\tau^2$
and the shift $s_1(\bar{\tau}^1)$ at time $\bar{\tau}^1$, then he
can obtain the shift $s_2(\tau^2)$ at time $\tau^2$.

(ii) If a user receives the trajectory $\bar{\tau}^2 =
\varphi_1(\tau^1)$ in the vicinity of time $\tau^1$ and the shift
$s_2(\bar{\tau}^2)$ at time $\bar{\tau}^2$, then he can obtain the
shift $s_1(\tau^1)$ at time $\tau^1$.

This interpretation of the constraint system has important
consequences. Let us define the {\it past echo functions}
$\varepsilon_i$ as follows:
\begin{equation}
\varepsilon_1 = \varphi_2 \circ \varphi_1 \, , \qquad \varepsilon_2
= \varphi_1 \circ \varphi_2 \, .
\end{equation}
These (past) echo functions have the following geometric
interpretation (see Fig. \ref{fig:echo}):
\begin{itemize}
\item[(i)]
If $\gamma_1$ receives at time $\tau^1$ a signal after being echoed
by $\gamma_2$, it must be emitted at time $\varepsilon_1(\tau^1)$.
\item[(ii)]
If $\gamma_2$ receives at time $\tau^2$ a signal after being echoed
by $\gamma_1$, it must be emitted at time $\varepsilon_2(\tau^2)$.
\end{itemize}

The proper time intervals $[\varepsilon_1(\tau^1), \tau^1]$ and
$[\varepsilon_2(\tau^2), \tau^2]$ are named (causal) echo intervals,
{\em i.e.}, an {\em echo interval} is the interval between
the emission of a signal by an emitter and its reception
after being reflected by the other emitter (see Fig.
\ref{fig:echo}).

\begin{figure*}[htb]
    \includegraphics[angle=0,width=0.77\textwidth]{%
    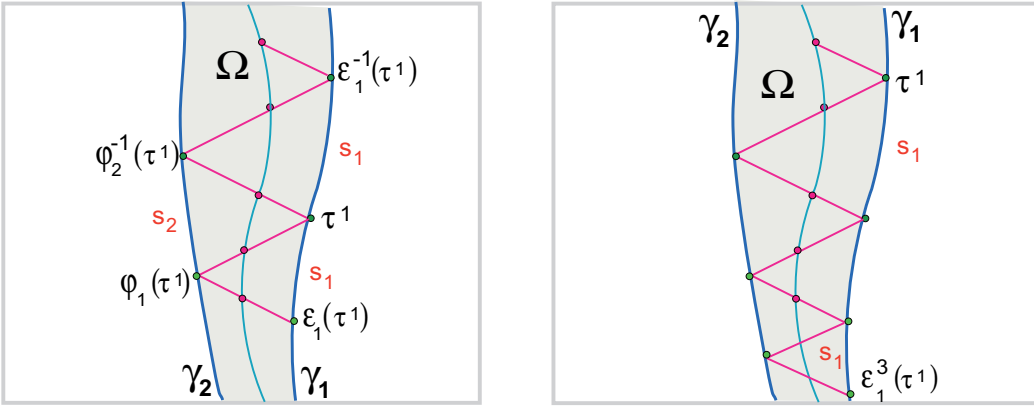}
    \caption{       \label{fig:master}
            Geometric interpretation of the master delay equation: (a) If a
            user receives the emitter positioning data $\{\tau^1, \tau^2;
            \bar{\tau}^1, \bar{\tau}^2\}$ along the arc $[\varphi_1
            (\tau^1),\varphi_2^{-1}(\tau^1)]$ and the emitter shift $s_1$ in the
            echo interval $[\varepsilon_1(\tau^1), \tau^1]$, then the user knows
            the shift $s_2$ in the arc $[\varphi_1
            (\tau^1),\varphi_2^{-1}(\tau^1)]$ as a consequence of
            (\ref{constraint-shift-1-bis}). Therefore the user knows the shift
            $s_1$ in the echo interval $[\tau^1 , \varepsilon_1^{-1}(\tau^1)]$
            as a consequence of (\ref{constraint-shift-2-bis}). (b) The delay
            master equations
            can be applied repeatedly in order to obtain an emitter shift
            further from an echo interval.}
\end{figure*}

Let us suppose that a user receives the emitter acceleration
$\alpha_1$ (and so he knows the shift $s_1$) in the echo interval
$[\varepsilon_1(\tau^1), \tau^1]$, and that he also receives the emitter
positioning data $\{\tau^1, \tau^2; \bar{\tau}^1, \bar{\tau}^2\}$
along the arc $[\varphi_1 (\tau^1),\varphi_2^{-1}(\tau^1)]$ that is,
he knows the emitter trajectories $\varphi_i(\tau^i)$ along this
arc. Then the user knows the shift $s_2$ along the arc $[\varphi_1
(\tau^1),\varphi_2^{-1}(\tau^1)]$ as a consequence of
(\ref{constraint-shift-1-bis}) (see Fig. \ref{fig:master}(a)).
Therefore the user knows the shift $s_1$ in the echo interval
$[\tau^1 , \varepsilon_1^{-1}(\tau^1)]$ as a consequence of
(\ref{constraint-shift-2-bis}). And so on (see Fig.
\ref{fig:master}(b)).

We can obtain the analytical expression of this fact by replacing
$\tau^2$ by $\varphi_1(\tau^1)$ in equation
(\ref{constraint-shift-1-bis}) and substituting in
(\ref{constraint-shift-2-bis}). Then we arrive to the {\it delay
master equation}:
\begin{equation} \label{master-1}
s_1(\tau^1) =
\frac{\dot{\varphi}_2(\varphi_1(\tau^1))}{\dot{\varphi}_1(\tau^1)}
s_1(\varepsilon_1(\tau^1)) \, .
\end{equation}
In a similar way, by replacing $\tau^1$ with $\varphi_2(\tau^2)$ in equation
(\ref{constraint-shift-2-bis}) and substituting in
(\ref{constraint-shift-1-bis}), we obtain:
\begin{equation} \label{master-2}
s_2(\tau^2) =
\frac{\dot{\varphi}_2(\tau^2)}{\dot{\varphi}_1(\varphi_2(\tau^2))}
s_2(\varepsilon_2(\tau^2)) \, .
\end{equation}

The delay master equations (\ref{master-1}) and (\ref{master-2}) can
be written in terms of {\it the echo operators} $Q_i(\tau^i)$ as:
\begin{eqnarray} \label{master-1b}
s_1(\tau^1) = Q_1(\tau^1) s_1(\varepsilon_1(\tau^1)), \ \
Q_1(\tau^1) \equiv
\frac{\dot{\varphi}_2(\varphi_1(\tau^1))}{\dot{\varphi}_1(\tau^1)},
\quad
\\[1mm]
\label{master-2b} s_2(\tau^2) = \frac{1}{Q_2(\tau^2)}
s_2(\varepsilon_2(\tau^2)), \ \ Q_2(\tau^2) \equiv
\frac{\dot{\varphi}_1(\varphi_2(\tau^2))}{\dot{\varphi}_2(\tau^2)}.
\quad
\end{eqnarray}

Evidently, we can obtain the emitter shifts further from an echo
interval by applying the delay master equation repeatedly. This fact
can be expressed by using {\it the n-echo operators} $Q_i^n(\tau^i)$
(see Fig. \ref{fig:master}(b)):
\begin{eqnarray} \label{n-master-1}
s_1(\tau^1) = Q_1^n(\tau^1) s_1(\varepsilon_1^{n}(\tau^1))  \, ,
\\
\label{n-master-2} s_2(\tau^2) = \frac{1}{Q_2^n(\tau^2)}
s_2(\varepsilon_2^{n}(\tau^2))\, ,
\\
Q_i^n(\tau^i)\! \equiv \! \prod_{r=0}^{n-1}
Q_i(\varepsilon_i^r(\tau^i)) \, .
\end{eqnarray}
These equations allow to state:\\[0.5mm]
{\it Statement 8}.-- A user may know the shift of an emitter along
his trajectory provided that he receives the shift during a sole
echo interval and the emitter positioning data along his
trajectory.

\subsection{Getting the dynamics by means of the delay master
equation} \label{subsec-VI-B}

Now, we can use the delay master equation to improve the results in Sec.
\ref{section-III}. Indeed, if we take into account these results and
statement 8, we arrive to:\\[0.5mm]
{\it Statement 9}.-- If a user receives the emitter positioning data
$\{\tau^1, \tau^2; \bar{\tau}^1, \bar{\tau}^2\}$ along his
trajectory and the acceleration of one of the emitters during a sole
echo interval, then this user can obtain a full information about
his dynamics and the dynamics of the emitters.

In order to obtain all this information in a specific situation it
is worth analyzing what is the minimum set of equations which are
necessary. We have obtained the master delay equations
(\ref{master-1}) and (\ref{master-2}) from the constraint system
(\ref{constraint-shift-1-bis})-(\ref{constraint-shift-2-bis}), and a
straightforward calculation allows to show:\\[0.5mm]
{\it Statement 10}.- If the emitter trajectories in the grid $\varphi_1(\tau^1)$
and $\varphi_2(\tau^2)$ are known, then one of the constraint
equations
(\ref{constraint-shift-1-bis})-(\ref{constraint-shift-2-bis}) and one
of the master delay equations (\ref{master-1})-(\ref{master-2}) imply
the full constraint system
(\ref{constraint-shift-1-bis})-(\ref{constraint-shift-2-bis}).

Then, we can slightly modify the steps given in subsections
\ref{subsection-III-D} and \ref{subsection-III-E} in order to obtain
all the system and user information from a minimal set of public
data.
\begin{description}
\item {\em Received user data:} the emitter positioning data
$\{\tau^1, \tau^2; \bar{\tau}^1, \bar{\tau}^2\}$ along the user
trajectory and the acceleration of an emitter, say $\alpha_1$, in an
echo interval.
\item[] Step s1: From the pairs $\{\tau^1; \bar{\tau}^2\}$
and $\{\tau^2; \bar{\tau}^1\}$, determine the emitter trajectory
functions $\varphi_1(\tau^1)$ and $\varphi_2(\tau^2)$, respectively.
\item[] Step s2$'$: From the pair $\{\tau^1; \alpha_1\}$,
determine the emitter acceleration scalar $\alpha_1(\tau^1)$ in
the echo interval.
\item[] Step s3$'$: From the acceleration scalar $\alpha_1(\tau^1)$
obtained in step s2$'$, determine the shift $s_1(\tau^1)$ with
respect to an inertial system $\{\texttt{u},\texttt{v}\}$ in the
echo interval.
\item[] Step s3$''$: From the shift $s_1(\tau^1)$ in the
echo interval obtained in step s3$'$, determine the shift
$s_1(\tau^1)$ with respect to an inertial system
$\{\texttt{u},\texttt{v}\}$ along the user trajectory:
    $$\qquad s_1(\tau^1) =
\frac{\dot{\varphi}_2(\varphi_1(\tau^1))}{\dot{\varphi}_1(\tau^1)}
s_1(\varepsilon_1(\tau^1)) \, , \quad \varepsilon_1 = \varphi_2
\circ \varphi_1 \, .$$
\item[] Steps s4-s7: From the function $\varphi_2(\tau^2)$ obtained
in step s1 and the shift $s_1(\tau^1)$ obtained in step s3$''$,
determine: the shift $s_2(\tau^2)$ with respect to the inertial
system $\{\texttt{u},\texttt{v}\}$ and the acceleration scalar
$\alpha_2(\tau^2)$, the metric function in emission coordinates, the
transformation from emission to inertial null coordinates
$\{\texttt{u},\texttt{v}\}$, and the proper time history of the
emitters in these inertial coordinates along the whole emitter world
lines.
\item[] Steps u1-u7: From the steps s1, s2, s3$''$, s4 and s6 and the pair
$\{\tau^1; \tau^2\}$, determine: the user trajectory in the grid,
the user world line in the inertial system
$\{\texttt{u},\texttt{v}\}$, the user proper time function $\tau =
{\cal T}(\texttt{u})$, the proper time history of the user in the
inertial null coordinates $\{\texttt{u},\texttt{v}\}$, the shift
$s(\tau)$ of the user with respect the inertial system
$\{\texttt{u},\texttt{v}\}$ and the user acceleration
$\alpha(\tau)$, and the proper time history of the user in emission
coordinates.
\end{description}

\subsection{The delay master equation in positioning with inertial
emitters} \label{subsec-VI-C}

Let us suppose that the user receives along his trajectory a set of
emitter positioning data $\{\tau^1, \tau^2; \bar{\tau}^1,
\bar{\tau}^2\}$ that leads, following step s1, to the emitter
trajectories (\ref{grid-inertial}) in the grid. Thus, the
echo function $\varepsilon_1$ and the echo operator $Q_1$ are,
respectively,
\begin{equation}
\varepsilon_1(\tau^1) = \lambda^2 \tau^1 + p \, , \qquad Q_1(\tau^1)
= 1 \, ,
\end{equation}
where $p \equiv \lambda \tau_0^2 + \tau_0^1$. Then, the delay master
equation for the shift $s_1(\tau^1)$ takes the expression:
\begin{equation} \label{master-inertial}
s_1(\tau^1) =  s_1(\lambda \tau^1 + p) \, .
\end{equation}

Let us suppose moreover that, following step s2$'$, the data
$\{\tau^1; \alpha_1\}$ determine that the acceleration scalar
identically vanishes in an echo interval,
$\alpha_1(\tau^1)=0$. Then, following step s3$'$, a null inertial
system $\{\texttt{u},\texttt{v}\}$ exists such that the shift in
this echo interval is $s_1(\tau^1) = 1$. Now, in step s3$''$,
we apply the delay master equation (\ref{master-inertial}) and
obtain $s_1(\tau^1) = 1$ along the user trajectory. At this point,
following the steps s4-s7 and u1-u7 we obtain all the system and
user information as we did in Sec. \ref{section-IV}.

\subsection{The delay master equation in positioning with stationary
emitters} \label{subsec-VI-D}

Let us suppose that the user receives along his trajectory a set of
emitter positioning data $\{\tau^1, \tau^2; \bar{\tau}^1,
\bar{\tau}^2\}$ that leads, following step s1, to the emitter
trajectories (\ref{grid-stationary}) in the grid. Thus, the
echo function $\varepsilon_1$ and the echo operator $Q_1$ are,
respectively,
\begin{equation}
\varepsilon_1(\tau^1) = \tau^1 - 2q \, , \qquad Q_1(\tau^1) =
\omega^2 \, .
\end{equation}
Then, the delay master equation for the shift $s_1(\tau^1)$ takes the
expression:
\begin{equation} \label{master-stationary}
s_1(\tau^1) =  \omega^2 s_1(\tau^1 - 2q) \, .
\end{equation}

Let us suppose moreover that, following step s2$'$, the data
$\{\tau^1; \alpha_1\}$ determine that the acceleration scalar takes
the constant value $\alpha_1(\tau^1)=\frac{1}{q} \ln \omega$ in an
echo interval. Then, following step s3$'$, a null inertial
system $\{\texttt{u},\texttt{v}\}$ exists such that the shift in
this echo interval is $s_1(\tau^1) = \exp(\alpha_1 \tau^1)$.
Now, in step s3$''$ we apply the master delay equation
(\ref{master-stationary}) and obtain $s_1(\tau^1) = \exp(\alpha_1
\tau^1)$ along the user trajectory. At this point, following the
steps s4-s7 and u1-u7 we obtain all the system and user information
as we did in Sec. \ref{section-V}.

\subsection{The delay equations for the emitter accelerations}
\label{subsec-VI-F}

In statement 7 we can replace the shifts $s_1$ and $s_2$ with the
accelerations $\alpha_1$ and $\alpha_2$ as a consequence of the
public data constraint equations (\ref{constraint-acceleration-1})
and (\ref{constraint-acceleration-2}). Then, from these equations or
from the delay master equations (\ref{master-1b})-(\ref{master-2b}),
we can obtain the delay equations for the emitter acceleration
scalars:
\begin{eqnarray} \label{master-1-a}
\alpha_1(\tau^1) = \frac{\dot{Q}_1(\tau^1)}{Q_1(\tau^1)} +
\alpha_1(\varepsilon_1(\tau^1)) \dot{\varepsilon}_1(\tau^1) \, ,
\\[1mm]
\label{master-2-a} \alpha_2(\tau^2) = -
\frac{\dot{Q}_2(\tau^2)}{Q_2(\tau^2)} +
\alpha_2(\varepsilon_2(\tau^2)) \dot{\varepsilon}_2(\tau^2) \, .
\end{eqnarray}

Moreover, we can also obtain a restriction on the emitter
accelerations further from an echo interval:
\begin{eqnarray} \label{n-master-1-a}
\alpha_1(\tau^1) = \frac{\dot{Q}_1^n(\tau^1)}{Q_1^n(\tau^1)} +
\alpha_1(\varepsilon_1(\tau^1)) \dot{\varepsilon}_1^n(\tau^1) \, ,
\\[1mm]
\label{n-master-2-a} \alpha_2(\tau^2) = -
\frac{\dot{Q}_2^n(\tau^2)}{Q_2^n(\tau^2)} +
\alpha_2(\varepsilon_2(\tau^2)) \dot{\varepsilon}_2^n(\tau^2) \, .
\end{eqnarray}
Thus, as a consequence of these equations we can replace in
statement 8 the emitter shift with the emitter acceleration.

The delay equations (\ref{master-1-a})-(\ref{master-2-a}) for the
emitter accelerations follow from the master equations
(\ref{master-1b})-(\ref{master-2b}) but they are not sufficient
conditions.

Thus, if we know the acceleration of an emitter in an echo
interval we must: firstly, obtain the shift and, secondly, apply the
master equation, as explained in steps presented in section
\ref{subsec-VI-B}. If, on the contrary, we first apply the delay
equation for the acceleration and, secondly, we determine the shift,
we could lost a part of the information that the master equation
provides.

We can better understand this point with an example. Let us suppose
that the user receives along his trajectory a set of emitter
positioning data that leads to the emitter trajectories
(\ref{grid-stationary}) in the grid. And let us also suppose that he
receives the acceleration of the emitter $\gamma_1$ in an
echo interval with a constant value $\alpha_1$. Then, we can
obtain the shift in this echo interval and the master
equation (which takes the expression (\ref{master-stationary}))
implies that, under a continuity assumption for the shifts, the
accelerations takes, necessarily, the constant value
$\alpha_1(\tau^1) =\frac{1}{q} \ln \omega$.

Nevertheless, if we apply first the delay equation for the
accelerations, $\alpha_1(\tau^1)= \alpha_1(\tau^1-2q)$, we obtain
$\alpha_1(\tau^1) =\alpha_1$ independently of the value of
$\alpha_1$. This apparent no constraint on $\alpha_1$ is deceptive:
if we apply the steps s1-s7 presented in section
\ref{subsection-III-D} for a value of the acceleration $\alpha_1
\not= \frac{1}{q} \ln \omega$, we arrive to an inconsistency.

\section{Discussion and work in progress \label{discussion}}

In this work we have analyzed the constraints on the data received
by a user of a relativistic positioning system, and how these data
can afford information on the dynamics of the user and of the
emitters. We have shown that the user can obtain his acceleration
and the acceleration of the emitters provided that he receives the
emitter positioning data along his trajectory and the acceleration
of only one of the emitters and only during a (causal) echo interval.

We have presented a protocol organized in steps which allows to
obtain, from the minimal set of data, all the system and user
information, namely, the acceleration of the emitters and of the
user, the transformation from the emission to inertial null
coordinates, and the proper time history of the emitters and of the
user in this inertial system.

Our study shows that the delay master equation plays an essential
role in the internal behavior of a positioning system built in a
flat two-dimensional space-time. A forthcoming work should deal with
looking for a similar constraint in a four-dimensional space-time and in
presence of a gravitational field.

In a future extension to the four-dimensional case of the
two-dimensional methods used here we should take into account the
role that the angle between pairs of arrival signals could play in
obtaining information on the metric tensor and on the positioning
system.

\begin{acknowledgments}
This work has been supported by the Spanish Mi\-nis\-terio de Ciencia e
Innovaci\'on, MICIN-FEDER project FIS2009-07705.
\end{acknowledgments}

\appendix

\section{Two-dimensional kinematics in null coordinates}
\label{ap-A}

In a null coordinate system $\{\tau^1,\tau^2\}$ the space-time
metric depends on a sole {\em metric function} $m$:
\begin{equation} \label{A-metric-taus}
ds^2 = m(\tau^1,\tau^2) d \tau^1 d\tau^2 \, .
\end{equation}

The proper time history of an observer $\gamma$ is:
\begin{equation} \label{A-observer}
\tau^1= \psi_1(\tau),  \qquad  \tau^2 = \psi_2(\tau) \, ,
\end{equation}
and its tangent vector is:
\[T(\tau) = (\dot{\psi_1}(\tau), \dot{\psi_2}(\tau)) \, ,\]
where a dot means derivative with respect proper time. The unit
condition for $T$ becomes:
\begin{equation} \label{A-unit}
m(\psi_1(\tau),\psi_2(\tau)) = \frac{1}{\dot{\psi_1}(\tau)
\dot{\psi_2}(\tau)} \, .
\end{equation}
This relation implies that when the unit tangent vector of an
observer is known in terms of his proper time, the metric on the
trajectory of this observer is also known.

The proper time parameterized trajectory (\ref{A-observer}) is
tantamount to a (geometric) trajectory $\tau^2 = F(\tau^1)$ and a
proper time function $\tau = \tau(\tau^1)$ related and restricted by
the unit condition. Indeed, from one of the expressions
(\ref{A-observer}) we can obtain the proper time of the observer
$\gamma$, say:
\[ \tau = \tau(\tau^1) = \psi_1^{-1}(\tau^1) \, .
\]
Then, the trajectory is given by:
\[\tau^2=F(\tau^1) = \psi_2(\psi_1^{-1}(\tau^1)) \, , \]
and, in terms of $\tau^2 = F(\tau^1)$ and $\tau = \tau(\tau^1)$, the
unit condition (\ref{A-unit}) becomes:
\begin{equation} \label{A-trajectory}
[\tau'(\tau^1)]^2 = m(\tau^1, F(\tau^1)) F'(\tau^1) \, .
\end{equation}
From equation (\ref{A-trajectory}) it follows: if the the metric
function is known, (i) there always exists a congruence of users
having a prescribed proper time function, and (ii) the geometric
trajectory of a observer determines his local unit of time.

The acceleration of the observer (\ref{A-observer}) in null
coordinates  $\{\tau^1,\tau^2\}$ takes the expression:
\begin{equation}
a(\tau) = \left(\ddot{\psi_1} + (\ln m),_1 \dot{\psi}_1^2\,, \;
\ddot{\psi_2} + (\ln m),_2 \dot{\psi}_2^2\right) \, ,
\end{equation}
and the {\it acceleration scalar} $\alpha(\tau) \equiv \pm
\sqrt{-a^2(\tau)}$ is:
\begin{equation} \label{B-ac-scalar}
\alpha(\tau) = \frac{\ddot{\psi_1}}{\dot{\psi}_1} + (\ln m),_1
\dot{\psi}_1 = - \frac{\ddot{\psi_2}}{\dot{\psi}_2} - (\ln m),_2
\dot{\psi}_2 \, .
\end{equation}

The {\it dynamic equation}, i.e. the equation for the world lines
with a known acceleration $\alpha$, and consequently the {\it
geodesic equation} (when $\alpha = 0$), can be written as two
coupled equations for the proper time functions $\psi_1(\tau)$ and
$\psi_2(\tau)$:
\begin{equation} \label{A-dynamic-eq}
\frac{\ddot{\psi_1}}{\dot{\psi}_1} + (\ln m),_1 \dot{\psi}_1 =
\alpha(\tau) \, , \qquad  m \dot{\psi}_1 \dot{\psi}_2 = 1
\end{equation}
In (\ref{A-dynamic-eq}) the metric function $m(\tau^1, \tau^2)$ is known, and $m$
stands for $m(\tau^1(\tau), \tau^2(\tau))$; therefore, it is a coupled system.

\subsection*{Dynamic equation in flat metric}
\label{subsection-A}

In a two-dimensional flat space-time the metric function $m$ in null
coordinates $\{\tau^1, \tau^2\}$ factorizes:
\[m(\tau^1, \tau^2) = u'(\tau^1) v'(\tau^2) \, ,\]
where $\texttt{u}=u(\tau^1)$ and $\texttt{v}=v(\tau^2)$ give the
transformation to an inertial coordinate system $\{\texttt{u},
\texttt{v}\}$.

As a consequence of this factorization, the dynamic equation
(\ref{A-dynamic-eq}) can be partially integrate and it becomes:
\begin{equation} \label{A-dynamic-eq-flat}
\dot{\psi}_1(\tau) u'(\psi_1(\tau)) = \frac{1}{\dot{\psi}_2(\tau)
v'(\psi_2(\tau))} = s(\tau) \, ,
\end{equation}
where the {\it shift parameter} $s(\tau)$ is defined as:
\begin{equation} \label{A-shift}
s(\tau) \equiv \exp\left(\int \alpha(\tau) d \tau\right) \, .
\end{equation}
Note that $s(\tau)$ is, actually, a shift parameter since it could
be obtained as:
\begin{equation} \label{A-shift-beta}
s(\tau) = \sqrt{\frac{1 + \beta(\tau)}{1- \beta(\tau)}}
\end{equation}
where $\beta(\tau)$ is the relative velocity between the given
observer and an inertial one. The hyperbolic angle between both
observers is $\, \phi(\tau) = \ln s(\tau)$.

\end{document}